# Coherent CVA and FVA with Liability Side Pricing of Derivatives

Lou Wujiang[1]
1st Draft, June 11, 2014, updated August 31, 2015[2].

**Abstract**

This article presents FVA and CVA of a bilateral derivative in a coherent manner, based on recent developments in fair value accounting and ISDA standards. We argue that a derivative liability, after primary risk factors being hedged, resembles in economics an issued variable funding note, and should be priced at the market rate of the issuer's debt. For the purpose of determining the fair value, the party on the liability side is economically neutral to make a deposit to the other party, which earns his current debt rate and effectively provides funding and hedging for the party holding the derivative asset. The newly derived partial differential equation for an option discounts the derivative's receivable part with counterparty's curve and payable part with own financing curve. The price difference from the counterparty risk free price, or total counterparty risk adjustment, is precisely defined by discounting the product of the risk free price and the credit spread at the local liability curve. Subsequently the adjustment can be broken into a default risk component – CVA and a funding component – FVA, consistent with a simple note's fair value treatment and in accordance with the usual understanding of a bond's credit spread consisting of a CDS spread and a basis. As for FVA, we define a cost - credit funding adjustment (CFA) and a benefit - debit funding adjustment (DFA), in parallel to CVA and DVA and attributed to counterparty's and own funding basis. This resolves a number of outstanding FVA debate issues, such as double counting, violation of the law of one price, misuse of cash flow discounting, and controversial hedging of own default risk. It also allows an integrated implementation strategy and reuse of existing CVA infrastructure.

## 1. Introduction

Some banks are moving ahead with FVA while it's far from being considered a settled issue among academia researchers, accountants, quants, traders, and regulatory stakeholders. Indeed, complexity surrounding FVA has only increased as the discussion has been deepening. KPMG (2013) for example highlights nine propositions toward understanding and implementation of FVA. While accounting definition of fair value,

---

[1] The views and opinions expressed herein are the views and opinions of the author, and do not reflect those of his employer and any of its affiliates. The author wishes to thank Yi Tang, Yadong Li for their helpful comments.

[2] Prior to this update, a later research into uncollateralized swaps and implementation by Monte Carlo simulation is available in ssrn: Liability-side pricing of non-CCP swap and coherent CVA and FVA computation by regression/simulation, August 8, 2015, abstract_id=2498770.

ISDA documentation of legal terms of derivative settlement and collateralization, investment theory and practice, and banks' treasury and derivative desk operation dominate recent discussion, a difficulty at the center is a lack of consensus on what economic principles shall apply when incorporating funding costs into bilateral derivative pricing.

Two streams of research and development can be seen in the existing literature on this subject: one is characterized by an ad-hoc valuation adjustment to the risk-free price made to take into account of costs of funding the derivatives in addition to CVA/DVA, and the other by carefully constructing a Black-Scholes-Merton micro-economy where funding of the derivatives and hedging of counterparty credit exposure are integrated with delta hedging to replicate the option and to obtain a PDE (Burgard and Kjaer 2011, 2013) that governs the eventual fair price and its resultant valuation adjustments. With the first stream, there is little discussion on whether the risk-free price when added with all adjustments is arbitrage-free. With regard to the second stream, the resulted PDE directly responds to hedge and funding assumptions, no all of which are economically reasonable. Hull and White (2012), for example, point out that the extra collateral spread (Piterbarg 2010) earned by the hedge party runs the risk of ignoring the collaterals' own credit risk. A PDE framework does not automatically result in compatibility with the notion of a risk neutral measure, easily testified by the widely observed violation of the law of one price.

This paper adds a new model to the second stream by proposing the liability side pricing principle for uncollateralized derivatives. The key idea is that the *fair* instrument, for the parties of a bilateral derivative trade to hedge their counterparty exposure and to fund the derivative, is to have the party on the liability side deposit cash. The party making deposit is economically neutral as the deposit will earn a coupon rate same as the market rate of the depositor's debt. The deposit made out of the derivative's netting set would offset the derivative exposure via ISDA's Set-off provision, avoiding the need of otherwise controversial counterparty hedge by means of dealing own CDS or bonds. This setup departs from existing literature and allows a new Black-Scholes-Merton (BSM) equation to be derived and the risk-neutral pricing formula to be revised with the risk-free discount rate being replaced by the liability-side's debt interest rate. The newly defined fair value adjustments -- CVA and FVA -- are coherent in that they are holistic results of decomposing the total valuation adjustment into default risk and funding risk, they do not create arbitrage opportunity, they observe the law of one price, and they can be implemented together.

**2. Pricing uncollateralized options under bilateral credit and funding risks**

Party B (a hypothetic bank) and party C (a customer) enter into an option trade with the bank dynamically hedging the option with underlying stock which is financed in the repo or security lending market. Both parties have access to a liquid corporate bond market, primary or secondary, exogenous to the simple option economy.

**2.1 Economics of a bilateral trade**

Let's start by conducting an economic experiment. When party C has a positive exposure to party B (the derivative is a receivable to C), C asks B to deposit an amount of cash by promising to pay interest at the market rate of B's same ranked debt. Party B is economically indifferent as he could raise the cash from the bond market at the same rate. From B's balance sheet (B/S) perspective, the new liability assumed by issuing in the market place offsets his deposit (on the asset side) so that there is no net balance sheet impact[3]. From cash flow perspective, B receives the market rate from C and pays out the same to bond investors. No new cash flow is generated.

From party C's perspective, her derivative asset is now financed by B. No other forms of financing, including from her own treasury, is necessary.

If B is holding a receivable, then C would make a deposit earning her market rate. To summarize we have obtained the following economic proposition.

**Proposition 1 (Market funding of uncollateralized derivative)**: Parties in an uncollateralized bilateral derivative trade implicitly grant each other a funding obligation at their respective cash market rates.

Now imagine if B defaults, C has two transactions with B: the derivative receivable, and the deposit as a payable. Under common law, the right to set-off would apply to determine the final balance of party C's settlement amount. If the deposit amount is same as the derivative close-out amount, the derivative contract and the deposit set off completely. The net balance is zero and there is no default settlement cash flow.

Economically, party C sees it as a market neutral hedging strategy against B's default risk.

Reciprocally, the defaulting party, party B, has no default settlement cash flow either, no default windfall or shortfall. Consequently, there is nothing to hedge upon its own default. C's proposal of a deposit from B at his market rate therefore serves as a market neutral hedging strategy for B as well as for C.

**Proposition 2 (Market neutral counterparty exposure hedge)**: Parties in an uncollateralized bilateral derivative trade can hedge credit exposure mutually by making cash deposit in the same amount as the derivative's close-out amount.

This proposition looks same as a fully collateralized derivative under bilateral CSA (Credit Support Annex) where the derivative's mark-to-market is fully collateralized. A critical difference exists that the collateralization under CSA is contractual, i.e., an integral part of the trade, while here it is non-contractual although economically justified and implied. This leads to different rates being applied to cash collateral: cash rate (e.g., federal fund rate) with CSA, cash debt rate without CSA. Moreover, the deposit is made out of the derivative netting set and is covered by the Set-off provision of ISDA 2002 Master Agreement or under common laws.

Combining these two propositions, the cash deposit serves both purposes of hedging counterparty credit exposure and financing the derivative. The total economic cost of hedging and funding is the market rate of the depositor's debt. This is the basic economics of a bilateral derivative trade.

**2.2 Replicating an uncollateralized option**

---

[3] It is possible that leverage ratio calculation would include the asset under Basel 3.

Following Lou (2013), the wealth equation of a long option economy from party B's perspective is

$$\pi_t = M_t + (1-\Gamma_t)(V_t - L_t - N_t - \Delta_t S_t + L^s),$$

where $S$ denotes the stock price, $V$ option fair price, and $1-\Gamma$ the joint survival indicator. $L_t$ is a voluntary deposit account set up under Proposition 1, $L_t = L_t^+ - L_t^-$, $L_t^+$ the cash amount deposited or posted by party C to B that pays C's cash rate $r_c(t)$, and $L_t^-$ is the cash collateral by B to C earning B's cash rate $r_b(t)$. $L^s=(1+h)\Delta S$ is stock lending account with haircut $h$ on $\Delta$ shares of stock. The stock short sale proceeds is deposited with the stock lender who pays rebate interest at the rate of $r_s$. $r - r_s$ is the stock borrowing cost.

$M_t$ is the cash (or bank) account that earns the risk free deposit rate r. $N_t$ is B's debt account that issues short term rolled debt at par rate $r_N(t)$, $r_N(t)>=r(t)$. The account could be secured by the remaining asset of the economy and could have recourse to the bank. In the worst case, this is the senior unsecured debt account.

Self financing equation including default settlement is written as

$$dM_t = rM_t dt$$
$$+ (1-\Gamma_t)[d\Delta_t(S_t + dS_t) - \Delta_t Sq dt + dN_t - r_N N_t dt + dL - r_c L^+ dt + r_b L^- dt - dL^s + r_s L^s dt]$$
$$+ d\Gamma_t[L^s - \Delta_t S_t + (V_t^+ - L_t^+)(R_c + X(1-R_c)) - (N_t + V_t^- - L_t^-)(1 - X(1-R_b))]$$

where $R$ is the recovery rate, $X$ is a random marker of the defaulting party, $X=1$ if party B defaults. We have applied pre-default market price $V(\tau_-)$ as the close-out amount.

Enforcing full collateralization under Proposition 2, $L_t = V_t, L_t^+ = V_t^+, L_t^- = V_t^-$, the differential wealth equation becomes,

$$d\pi_t - r\pi_t dt = (1-\Gamma_t)[dV_t - rV_t dt - \Delta_t(dS_t - (r-q)S_t dt) - (r_N - r)N_t dt$$
$$- (r_c - r)V^+ dt + (r_b - r)V^- dt + (r_s - r)L^s dt] + d\Gamma_t X \overline{R_b} N_t$$

where $\overline{R_b} = 1 - R_b$, the loss given default for the debt account $N_t$.

Setting $\pi_t=0$ leads to $N_t=h\Delta_t S_t$, showing that the economy only needs to finance the residual stock lending margin.

Assuming zero haircut, then $N_t=0$, the jump term disappears from the portfolio equation. Now apply Ito's lemma, assume delta hedge under the usual geometric Brownian motion stock price ($dS=\mu S dt + \sigma S dW$), and set $dt$ term to zero, we obtain the following PDE for the fair price of a bilateral option:

$$\frac{\partial V}{\partial t} + (r_s - q)S\frac{\partial V}{\partial S} + \tfrac{1}{2}\sigma^2 S^2 \frac{\partial^2 V}{\partial S^2} + r_b V^- - r_c V^+ = 0$$

## 2.3 Liability side pricing principle

For a pure derivative asset or receivable, $V^-=0$, $r_b$ term drops out and the fair price $V$ is governed by $r_c$, the cash curve of party C who is on the liability or payable side. A derivative asset's fair value therefore does not depend on an investor's cost of fund. This supports Hull and White (2012)'s argument for adherence to the law of one price.

For a pure derivative liability or payable, however, party C's funding curve drops out, and $V$ is governed by its own curve $r_b$. This is nothing unusual, as a liability is priced off its issuer, rather than its buyer.

Whether party B is holding a receivable or payable, the bilateral option is always priced on its liability side, same as a note being priced by the note issuer's credit and liquidity.

We now have counterparty's funding rates accompanying the positive and negative parts of the derivative fair value, replacing the usual $rV$ term in the BSM equation. An intuitive explanation is that with non-defaultable counterparties, derivative fair value accrues at the risk free rate; with defaultable counterparties, however, the positive part (receivable) will accrue at the counterparty's cash rate while the negative part at own cash rate. This can be stated as follows.

**Proposition 3 (Principle of Liability Side Pricing)**: A primary risk factor stripped bilateral derivative prices the derivative exposure at the market rate of the liability side's debt.

The bilateral option PDE can be derived in a simpler way with a direct application of the liability side pricing principle. Following Hull and White's textbook approach, we write hedge portfolio and finance equations

$$\pi_t = V_t - \Delta_t S_t,$$
$$d\pi_t = dV_t - \Delta_t dS_t = (r_c V^+ - r_b V^-)dt - (r_s - q)\Delta_t S_t dt$$

The right hand side of the portfolio differential equation consists of stock financing cost at the repo rate, and derivative financing at rates dictated by the liability side pricing principle. Applying Ito's lemma and delta hedge, setting $dt$ term to zero results in the same PDE.

The requirement that the primary risks be stripped makes the derivative a plain debt, albeit with variable principal exposure. A structured note linked to equity, commodity, or foreign exchange rates can be seen as a hybrid debt that should price at the issuer's credit rate once the underlying risk is hedged.

In the classic BSM theory, a delta hedged option performs like a risk free portfolio. With defaultable counterparties, the same portfolio becomes a credit risky debt of the party on the liability side. This principle, therefore to some extent, is an extension of the risk neutral pricing theory to uncollateralized derivatives. In fact, applying Feynman-Kac theorem straightforwardly leads to the same discounted payoff expectation formulae, except that the risk free discount rate is replaced by the debt rate that sticks to the liability side. The equivalent martingale measure for the underlying stock price remains the same.

**Proposition 4 (Extended Risk-neutral Pricing Formula)**: A bilateral derivative's no-arbitrage price is the expected risky discount of the derivative payoff under the risk neutral measure Q, with filtration sufficiently enlarged to include the counterparty's credit and funding rates,

$$V(t) = E_t^Q [e^{-\int_t^T r_e du} V(T)],$$
$$r_e = r_b I(V(u) \leq 0) + r_c I(V(u) > 0).$$

where $V(T)$ is a $T$-filtration measureable random variable. In general risky discount factor $D(t) = e^{-\int_0^t r_e du}$ depends on the local fair value so the expectation is coupled or recursive.

### 2.4 Bilateral option PDE with non-zero haircut

Haircut, as an overcollateralization measure to reduce stock lender's risk to volatility, has not been taken up adequately in the FVA literature. Basel 3 stipulates that the standard supervisory haircut for a non-main-index equity is 25%. Under the liability side pricing setup, the residual stock market value, i.e., the haircut amount, needs to be financed by the bank's treasury. Theoretically, the debt account $N_t$ could have a subordinated claim on additional stock market value when financing is closed[4]. In an early paper, Lou (2013) described this feature as endogenous recovery. The applicable LGD on $N_t$ is therefore zero. Accordingly the jump term in the differential portfolio wealth equation disappears. So again there will be no default settlement cash flow. As $N_t$ has no credit loss risk, the applicable rate spread $r_N - r$ should not contain compensation for party B's own default risk. It could be zero or up to the funding or liquidity cost or basis.

Below is a PDE with haircut and a collateral account that pays $r_L$ on posted amount, where $L$ is the amount of collateral posted by C under a weak CSA,

$$\frac{\partial V}{\partial t} + (r-q)S\frac{\partial V}{\partial S} + \tfrac{1}{2}\sigma^2 S^2 \frac{\partial^2 V}{\partial S^2} - rV - \gamma S\frac{\partial V}{\partial S} + (r_b - r)(V-L)^- - (r_c - r)(V-L)^+ - L(r_L - r) = 0,$$
$$\gamma = (r_N - r)h + (1+h)(r - r_s).$$

The last term reflects cost of funding the collateral under CSA. Obviously, with full collateralization, the derivative funding cost in the PDE returns to the risk-free rate, confirming discounting with the OIS curve.

### 2.5 Comparison of PDE with earlier results

---

[4] Or the additional cash posted to the stock lender, when short a stock.

There have been a number of efforts in extending the Black-Scholes-Merton PDE to study the funding cost impact on derivatives pricing. Lou (2010) recognizes funding asymmetry in that deposit and borrowing of cash never earn the same rate. Piterbarg (2010) builds an extra collateral return into the BSM equation. Burgard and Kjaer (2011, 2013) consider funding costs in a bilateral counterparty credit risk setting. Lou (2013) discusses realistic stock financing cost while also covering bilateral default risk, finding that own default risk could be mitigated if the derivative is segregated, which entails an endogenous par recovery.

These and other similar researches build an asset buyer's funding cost into valuation of a derivative asset and arguably belong to private valuation, rather than fair value (Hull and White 2014.) As we should see, private values are just as important as fair value, for they determine the trading desk's bid and ask prices.

The extended bilateral option PDE with zero haircut stock financing can be generalized in the form below.

$$\frac{\partial V}{\partial t} + (r_s - q)S\frac{\partial V}{\partial S} + \frac{1}{2}\sigma^2 S^2 \frac{\partial^2 V}{\partial^2 S} - rV + f_b(V-L)^- - f_c(V-L)^+ - L(r_L - r) - f_N N = 0,$$

The last four terms are adjustments, including benefit on own default and funding, cost on counterparty's default and funding, extra cost paid to get cash collateral, and a charge from the treasury as $N$ is the funding amount from the bank's treasury.

Hull and White (2014) have the simplest case, with $h=0$, $L=0$, $N=0$, and $f_c=r_d - r$, where $r_d$ is a derivative financing rate, and $f_b=0$. $r_d$ is more of a placeholder and does not associate with any counterparty. Piterbarg (2011) only considers collateral spread, a special case by setting $f_N=f_b=f_c=0$.

Burgard and Kjaer (2011) initially uses CDS to hedge both parties' credit exposure, so $f_b$ and $f_c$ are CDS implied credit spread, specifically, $f_b=\lambda_b R_b$, $f_c=\lambda_c R_c$ where $\lambda_b$ and $\lambda_c$ are the usual default intensities. The default intensity notion is subsequently dropped and replaced with bond yield as using a zero-coupon-zero-recovery bond as a counterparty risk hedge instrument is favored over the controversial CDS. Burgard and Kjaer (2013) employ different strategies to deal with the hedging error, $N_t$ term in our notation. Each strategy would result in variations of the coefficients in the PDE.

In our case, $N=0$, $f_b=r_b - r$ and $f_c=r_c - r$. Crepey (2013) and Brigo et al (2014) have also shown similar PDEs.

Variations in PDEs are owing to different funding and hedging arrangements, whether derived from a replication argument or from a risk neutral approach. Our derivation makes no reference to risk neutral pricing theory, unlike Lou (2013) and Brigo et al (2014). The connection to the risk neutral world, however, is not lost. What the liability side pricing principle says is that, after the primary risk factors are hedged, a bilateral derivative becomes a watered-down debt instrument, similar to a variable notional funding note, which shall be priced at the issuer's current market levels. This approach is a natural extension of the risk neutral pricing principle for the purpose of pricing bilateral defaultable derivatives.

The advantage of a risk-neutral based derivation is that it avoids the controversial issues of hedging own default risk, by selling own CDS protection or dealing in own debts, the former illegal and the latter imprudent and impractical. A close examination of the micro financing structure of a standalone option economy reveals that it is never possible for the economy to buy back its debt (Lou 2013). In fact in such a segregated derivative economy, there is no own default benefit to hedge. Here we essentially let the exposed side (weather the bank or its customer) hedge its exposure by means of shorting the other party's bond with the other party as the security lender, a more familiar but equivalent strategy comparing to the par deposit which has nil impact to the balance sheet.

Solely for the purpose of fair value measurement, there is no need to consider hedging one's own default risk. IFRS 13 stipulates an exit price of a would-be liability transfer to a third party of the same credit quality. To assume the hedged derivative liability, the 3$^{rd}$ party would charge its senior unsecured rate so that the PDE derived will again show the inclusion of the liability side's funding curve as presented above.

With regards to hedging of counterparty default risk, the use of bond as a hedge instrument and a continuously rolled par CDS will result in slightly different rate applied to the exposure to counterparty ($V^+$): the former leads to the cash bond rate for example Lou (2013), and the latter the synthetic funding rate. Economically, hedging with bonds is fully funded, but buying CDS protection is unfunded and relies on the performance of the protection seller at the time of the obligator default. Recognizing this important difference is the key to develop a coherent view of CVA and FVA that CVA is the unfunded adjustment while the sum of CVA and FVA becomes the funded, full adjustment.

## 3. Fair Value Adjustments

As valuation adjustments are components of the accounting fair value measurement, it's helpful to review the fair value accounting treatment of a simple note receivable (or payable).

### 3.1 Valuation adjustments example – simple note receivable

Party C holds a note receivable issued by party B. If the price of the note, denoted by $V$, is directly observable from the market place, she simply records $V$ on the asset side of her balance sheet. Or she could book an asset at the risk-free price $V^*$, then subtracting a credit valuation adjustment (CVA) with a contra account.

CVA is defined to reflect the default risk of the issuer of the note, typically measured in terms of the CDS curve. The implied bond price from a traded CDS curve, however, seldom agrees with the cash bond price. The difference that needs to be booked as yet another fair value adjustment reflects other funding factors such as bond and CDS market liquidity disparity, tax, and accounting (Longstaff et al 2005), and this becomes the base of a funding valuation adjustment (FVA). For the issuer, the credit component and the liquidity or basis component add up to his total financing cost.

Take for example, for a T maturity, zero coupon, zero recovery bond, let's denote $V$, $V^*$, $V\tilde{\,}$ the market (cash) price, risk-free price, and the CDS implied (risky) price. Assuming flat and deterministic continuously compounding yield $y$, rate $r$, and default intensity $\lambda$, we have the following,

$$V_t = e^{-y(T-t)},$$
$$V_t^* = e^{-r(T-t)},$$
$$V_t^\sim = e^{-(r+\lambda)(T-t)}.$$

For party C, a buyer of B's bond, $CVA = V^* - V^\sim$, and $CFA = V^\sim - V$, so that $V^* - CVA - CFA = V$.

Symmetrically for party B, the issuer, $DVA = V^* - V^\sim$, and the debit funding adjustment (DFA) is defined as $DFA = V^\sim - V$, so that $-V^* + DVA + DFA = -V$.

It is important to see from this simple and agreeable example that CVA and FVA are articles introduced to accommodate our way of analysis of the counterparty risk and funding factors, credit for default risk and everything else under the funding basis.

These prices ($V, V^*, V^\sim$) are solutions to the simple ordinary differential equation: $\frac{dV}{dt} - r_F V = 0$, when the funding curve $r_F$ is specified as the risk-free curve $r$, the synthetic funding curve $r^\sim$, and the cash curve y, respectively.

Assuming a deterministic yield and taking out of the stock price, the bilateral option PDE in section 2.2 reduces to $\frac{dV}{dt} - r_F V = 0$, where the funding curve $r_F$ is specified as $r_b$ for B's debt or $r_c$ for C's debt, respectively. Zero coupon bond pricing is reproduced.

The simple receivable/payable note is of course not a usual derivative instrument, but a bond-like structured note can be created easily. The point is that a bilateral derivative pricing model should be able to price the cash bond as a special case, i.e., a benchmark test case (Morini and Prampolini 2011).

**3.2 Multi-curve setting**

To facilitate valuation adjustment definition, let's introduce two sets of funding curves and their resultant prices, in addition to the usual risk free curve $r(t)$ and risk free price $V^*$.

**Synthetic Funding Curve**: A firm's synthetic (funding) curve is a CDS implied funding curve, with its short rate denoted by $r^\sim$. It has a non-negative spread curve over the risk free curve, i.e., $r^\sim - r \geq 0$. Under zero-recovery assumption, this corresponds to $r+\lambda$, where $\lambda$ is the default intensity in the reduced form modeling approach.

The risky price $V^\sim$ of a bilateral defaultable derivative is a price solved from the PDE when each counterparty's funding curve is assumed to be at its synthetic curve. This allows separation of the credit risk component of the financing cost so that it can be risk managed in the CDS market.

**Cash Funding Curve**: A firm's cash (funding) curve is the financing curve observed in the debt capital market or secondary bond market, typically for senior unsecured ranked debts. Let $r_b, r_c$ denote the cash (short) rates of party B and party C.

The fair price *V* of a derivative is when both credit and liquidity and other basis risks are considered. This is the price that goes into the firm's books and records.

These prices are all solutions to the same bilateral option PDE when appropriate curves are placed. For example, to solve for the risky price $\tilde{V}$, one simply replaces the cash rates $r_b$ and $r_c$ with the synthetic rates.

Other curves, such as the repo curve and collateral curve, may or may not enter the fair value PDE depending on whether they are market observable.

**3.3 Total counterparty risk adjustment (CRA)**

As the fair value *V* solved from the extended PDE fully incorporates counterparty credit risk and derivative funding cost, the total counterparty risk adjustment (CRA) is trivially the difference of *V* and the (counterparty) risk-free value $V^*$. Let $U = CRA = V^* - V$. Subtracting the BSM equation for $V^*$ from the extended PDE leads to,

$$\frac{\partial U}{\partial t} + (r_s - q)S\frac{\partial U}{\partial S} + \tfrac{1}{2}\sigma^2 S^2 \frac{\partial^2 U}{\partial S^2} - r_e U + (r_e - r)V^* = 0$$

Noting $U_T=0$, application of Feynman-Kac theorem immediately leads to the CRA formula precisely,

$$U = E_t[\int_t^T (r_e - r)V^*(s)e^{-\int_t^s r_e du}ds]$$

Intuitively, the total adjustment made to the risk-free price is the sum of liability side discounted excess return ($r_e$-$r$) on a notional amount of $V^*$.

**3.4 Coherent CVA and FVA decomposition**

Adjustment due to counterparty credit risk (CVA) was well understood in early 1990's (Tang and Li). CVA definitions depend on CDS, which unfortunately has become much less liquid following the financial crisis and increasingly unlikely to be available for non-advanced markets. CRA therefore offers an advantage as it naturally links to the bond spread, not the CDS spread. In advanced economies where there are liquid CDS markets, CRA can be further decomposed into a credit risk component – CVA – and a funding risk component – FVA, in parallel to the spread decomposition.

***3.4.1* CVA and CFA**. Consider from party B's perspective, for a pure receivable, $V \geq 0$, $r_b$ term drops out from the PDE, so own funding curve has no impact. CVA is defined as the difference of the risk free price and the synthetic price due to the counterparty's CDS implied funding, $CVA = V^* - \tilde{V}$.

Credit funding adjustment (CFA) is defined as the difference of the synthetic price and the cash price of the derivative, $CFA=\tilde{V}-V$. Total valuation adjustment $V^*-V=CVA + CFA$.

Here CVA is positive, and CFA is assumed positive and therefore understood as a cost[5],

$$U = CVA + CFA = E_t[\int_t^T (r_c - r)V^*(s)e^{-\int_t^s r_c du} ds]$$

This formula bears an intuitive explanation: the total adjustment made to the risk free price is the sum of liability side discounted excess return ($r_c$-$r$) on a notional amount same as the risk free price $V^*$. One can easily verify in the benchmark case of a simple note receivable.

To obtain CVA alone, one could replace the cash rate above with the synthetic rate. Note this formula is same as typically presented, for instance, in Tang and Li (2007) and Gregory (2010) for a zero recovery counterparty. For a portfolio, the risk-free value $V^*$ would be replaced by the expected positive exposure (EPE) to arrive at the counterparty's CVA, so a typical CVA implementation can be modified trivially to work for CFA or they can be computed simultaneously.

**3.4.2 DVA and DFA.** Similarly, for a pure payable to party B, $V \leq 0$, $r_c$ term drops out from the PDE so counterparty's funding curve has no impact and there is no CVA and CFA. Debit valuation adjustment (DVA) is defined as the difference between the price due to own synthetic funding curve and the risk free price, $DVA = \tilde{V} - V^* = |V^*| - |\tilde{V}|$.

Debit funding adjustment (DFA) is defined as the difference of the cash price and the synthetic price of the derivative, i.e., $DFA = V - \tilde{V} = |\tilde{V}| - |V|$ and, $V = V^* + DVA + DFA$. Here both DVA and DFA are benefit and can be computed similarly to CVA/CFA.

**3.4.3 Symmetry of CVA/DVA and CFA/DFA.** By flipping the PDE with a negative sign, one easily sees a party's CVA (CFA) is the other party's DVA (DFA).

**3.4.4** In general, if a derivative, for instance, an option trading strategy or an ATM swap, is a switcher, neither pure payable nor pure receivable during its life cycle, and if the two parties' financing curves are different, the derivative has to be priced by splitting into an asset part and a liability part. Specifically, the PDE asserts that the asset part is discounted by the counterparty's rate and the liability part by the pricing party's own rate, i.e., the liability side rules. We can always write $V = V^*$- $CVA + DVA - CFA + DFA$.

To strictly attribute CVA to counterparty default risk, leaving DVA to own default risk, we apply party C's synthetic curve while keeping own curve at the risk free curve to get a new price. The difference from the risk free price becomes the CVA. DVA, CFA and DFA can all be obtained with such an incremental curve shift scheme.

---

[5] The basis between CDS and cash bond is not guaranteed to be negative or positive as the bond spread is usually measured against LIBOR as an industry practice. Post-crisis, the CDS/Bond basis for major financial intermediaries has been significant, in the range of 30 to 50 bps. In this article, the funding basis is measured against the risk free rate, or OIS. Consequently, it is positive, if we accept that nothing can be more liquid than cash itself.

Purely for notational convenience, if we denote $P(f_b, f_c)$ a solution to the PDE, a pricing function of the risk-free curve, $f_b$ B's funding curve choice, $f_c$ C's funding curve choice, then $V^*=P(r,r)$, $\tilde{V}=P(\tilde{r}_b,\tilde{r}_c)$, and $V=P(r_b,r_c)$. Bilateral $bCVA = V^* - \tilde{V}$, and $bFVA = \tilde{V} - V$.

A presentation of costs -- CVA and CFA, and benefits -- DVA and DFA, can be done to make whole the incremental adjustments, for example,

$$CVA = P(r,r) - P(r,\tilde{r}_c),$$
$$DVA = -P(r,\tilde{r}_c) + P(\tilde{r}_b,\tilde{r}_c),$$
$$CFA = P(\tilde{r}_b,\tilde{r}_c) - P(\tilde{r}_b,r_c),$$
$$DFA = -P(\tilde{r}_b,r_c) + P(r_b,r_c),$$
$$V = V^* - CVA + DVA - CFA + DFA.$$

FVA therefore is a natural companion of CVA, in both concept and formulation. CVA has become quite a phenomenon, having dedicated methodologies, systems, business units and risk organizations, and regulatory capital charge. It would be formidable if FVA, a part of the adjustment to reflect derivative funding cost, will be undertaking the same path. Fortunately the PDE presented above allows a coherent definition of FVA along with CVA. One could reserve the term FVA for the total valuation adjustment, and split it into CVA and CFA. Or one can retain the CVA acronym to represent the total adjustment by changing its literals from *credit* valuation adjustment to *counterparty* valuation adjustment. With minimal work, existing CVA framework can easily accommodate FVA and capture CRA.

**3.5 Valuation adjustments with risk-free close-out**

To close this section, we list below the PDE when the default settlement rule follows the 1992 ISDA where the risk free price is regularly used.

$$\frac{\partial V}{\partial t} + (r_s - q)S\frac{\partial V}{\partial S} + \frac{1}{2}\sigma^2 S^2 \frac{\partial^2 V}{\partial^2 S} - (r + \lambda_b + \lambda_c)(V - V^*)$$
$$+ r_b(V^* - L)^- - r_c(V^* - L)^+ - r_L L = 0,$$

where $\lambda_b$, $\lambda_c$ are default intensities of party B and C respectively. For simplicity, the haircut has been set to zero above.

If we denote $U = V^* - V$ as the total valuation adjustment, the PDE becomes,

$$\frac{\partial U}{\partial t} + (r_s - q)S\frac{\partial U}{\partial S} + \frac{1}{2}\sigma^2 S^2 \frac{\partial^2 U}{\partial^2 S} - (r + \lambda_b + \lambda_c)U - (r_b - r)(V^* - L)^- + (r_c - r)(V^* - L)^+ + (r_L - r)L = 0,$$

Apply Feynman-Kac theorem to get the solution,

$$U(t) = \frac{1}{\beta_t Q_t} E \int_t^T \beta_u Q_u [-(r_b - r)(V^* - L)^- + (r_c - r)(V^* - L)^+ + (r_L - r)L] du$$

where $Q$ is the joint survival probability, $\beta$ is the risk free discount factor. The first term can split into DVA and DFA, second term CVA and CFA, and the third and last term is for collateral cost.

For ISDA 1992 parties who have yet to adopt the 2009 ISDA Close-out protocol, the above formulae can be used. A critical difference resulted from the risk-free price settlement is that the total valuation adjustment now has combined the default intensities and will involve incorrectly the joint survival probability, even for a pure asset, a concern examined by Brigo and Morini (2011).

### 3.5 Zero recovery CVA and FVA

Under zero recovery rate assumption, $\tilde{r}_c - r = \lambda_c$, and $r_c - r = \lambda_c + \eta_c$, where $\lambda_c$ is the default intensity and $\eta_c$ is the funding basis of party C, the coherent CVA and FVA are the following,

$$CVA = E_t[\int_t^T \lambda_c V^*(s) e^{-\int_t^s (r + \lambda_c) du} ds],$$

$$FVA = E_t[\int_t^T (\lambda_c + \eta_c) V^*(s) e^{-\int_t^s (r + \lambda_c + \eta_c) du} ds] - CVA.$$

Obviously CVA is same as usual definition of unilateral CVA which only considers counterparty default risk, i.e., no joint survival probability. Corresponding results under the risk-free settlement are given below,

$$CVA = E_t[\int_t^T \lambda_c V^*(s) e^{-\int_t^s (r + \lambda_c + \lambda_b) du} ds],$$

$$FVA = E_t[\int_t^T \eta_c V^*(s) e^{-\int_t^s (r + \lambda_c + \lambda_b) du} ds].$$

Note here both CVA and FVA involve B's intensity in addition to C's, i.e., which is economically counter-intuitive given the concerned derivative is held as a pure asset and is a liability of party C.

### 4. FVA Moving Forward – Discussions and Results

Unlike CVA which has well understood economics and originally unambiguous definition[6], FVA has been presented in many different ways and its economics vigorously debated among researchers from academia and the financial industry. In this section, we attempt to address a number of topics in the context of an uncollateralized option trading strategy between two defaultable counterparties.

---

[6] Traditional CVA definition is a result of the risk-free replacement value approach which, although allowing nice separation of the credit exposure from credit risk factors, has contributed to confusion and awkward economics (Brigo et al 2014). Since the adoption of ISDA 2002 Master Agreement and 2009 Close-out Amount Protocol has become the mainstream, such a definition needs to be revisited.

**4.1 FVA as component of accounting fair value measure**

Financing is a general financial term and funding cost is everywhere, in every financial product and every life cycle of a trade. The very first question is whether all funding costs shall affect or enter into fair value. FVA literally denotes a funding related valuation adjustment made to fair value accounting of a derivative. For any other funding costs that do not impact fair value, the firm or its treasury could calculate and exert a charge to the desk. We coin the term treasury funding charge (TFC) to distinguish these costs from FVA.

Simply put, FVA affects a bank's balance sheet (B/S) and income statement. TFC as a charge to the desk but a credit to the treasury, does not appear in the firm's consolidated B/S. A desk may maintain private bid and offer values for its derivatives which could have incorporated TFC, but it is fair value that appears in the firm's books and records.

With funding costs being divided between FVA and the funding charge, the next question becomes what goes where. By virtue of IFRS 13, a funding source and its cost typically observable in the market place shall enter into fair value. Repo financing of major stocks, for example, reflects the actual and marketable stock financing cost of dynamic hedging, so the impact to option fair value is indeed "market based measurement." On the other hand, private arranged stock financing (e.g., in the form of a secured loan) shall not count as a market observable funding source and its impact shall be evaluated by means of TFC and enter into bid/ask prices in due course.

What has been proven difficult to categorize is the funding cost of uncollateralized bilateral derivatives. The industry, supported by the latest IFRS updates (EY 2014), seems to have gained convergence on the need of taking funding cost into fair value, but whose funding cost is still wide open. Some argue and use dealers' own funding costs while other favor an industry average, including or excluding the pricing dealer itself. Hull and White reject FVA (2012, 2014) in that incorporating a bank's own funding cost in derivatives pricing would violate the law of one price, create arbitrage opportunities, and be at fault with investment theories.

The coherent CVA and FVA settles this question at the liability side's debt financing cost, which is market observable. In essence, without knowing which agent will be stepping in in an exit exercise assuming existence of a novation market, the only reasonable assumption to make is that agents of different funding costs will compete for the novation trade and at equilibrium they will settle at the level of the derivative liability party's cost.

The third question is whether FVA would violate the law of one price. While this is an issue with most, if not all, existing FVA definitions, it does not have to be the case. Such a law is derived for and only applies to investment in an *asset*. It is easy to see that assumption of a liability would demand different prices per the firm which takes up the liability. Our PDE would result in a unique price dependent on the liability side's funding curve, if the derivative is a pure receivable or asset to the pricing party. If the derivative is a pure payable (liability), then it depends on the liability issuing party's funding curve, in the same way a corporate bond's price would depend on its issuer's credit. Of course, when the derivative is a switcher of asset and liability, the price could depend on both party's funding curves. For IFRS fair value measurement purposes, the third party assuming the liability should be of the same credit quality of the current liability bearer, so our definition of FVA is IFRS compliant and there is no violation of law of one price.

A related issue is the overlapped accounting of FVA and DVA. For a derivative receivable, computing FVA on its mark-to-market corresponds to funding at the senior unsecured rate. The same funding, if raised from the market, would incur a DVA or DVA2 (Hull and White 2012) that would offset the component of FVA that is due to the dealer's own credit risk. The offset is complete only if there is no funding basis. Our definition of FVA solely attributes to funding or liquidity basis, on top of, rather than inclusive of, the default risk component. Also for the receivable, it is the other party's (rather than own) basis at work. By design, there is no overlap or double counting of FVA and DVA.

For a derivative payable or liability, DVA is now a standard accounting item, which Hull and White (2012) refer to as DVA1. An FVA benefit (or FBA), calculated on its MTM and at own funding rate, would overlap with DVA1. In the coherent CVA and FVA definition, the FVA benefit or DFA is attributed to the funding basis only and DVA to default risk, so again by design this is not an issue types. Coherent CVA and FVA as an accounting fair value measure therefore does not exhibit double counting issue.

**4.2 Treasury funding charge (TFC)**

Uncollateralized derivatives do not have their own financing markets and will have to rely on a bank's treasury function for funding. Funding operation linked to derivatives, however, is unconventional. The funding amount fluctuates often unpredictably, resulting in unmanageable revolving draw and pay-down. Funding duration changes and nominal maturity is subject to a number of early termination events. Some derivatives are effectively funding facilities in their own right, with complex structural features that would faint a traditional treasury. It is highly desirable for the treasury to have an advanced toolset to manage derivatives funding and liquidity risk.

In sections above, we show some contributing sources to FVA, such as counterparty exposure hedges in a bilateral derivative trade and repo funding of continuous stock trading in an option's dynamic replication process. Below we give examples where the effect of funding cost is best to be handled as an internal funding charge exerted by a bank's treasury on each trading desk.

**Example 4.2.1.** A treasury finances a desk's repo line with a counterparty. The desk borrows from treasury with a fixed term (e.g. 2 years) and principal amount (e.g. 100 mm) at LIBOR + 40 basis points, while it lends out at LIBOR + 100 bps. The bank may account the repo on an accrual basis. Then there is no fair value involved, nor is there an FVA. The trade does involve a funding cost (+ 40 bps) which is internal between the desk and the treasury, i.e., charge to the desk is gain to the treasury and from the bank's perspective, there is no net impact to its book and records.

**Example 4.2.2.** Same as above, but the desk is able to source the term fund from an external bank at LIBOR + 30 bps and has to book it as a separate trade. There is no internal funding charge here as no bank fund is used. The desk may book back-to-back repos and realize a funding arbitrage P&L on a present value basis, probably with marginal FVA to take into consideration of counterparty credit risk hedges.

**Example 4.2.3.** Similar to example 4.2.1., but the trade is done in a TRS format, fair valued and mark-to-market daily. As a result, the funding amount will float in accordance with the market value of the asset funded. A treasury could treat this as a flat funding principal and offer to fund the variation at its overnight rate. If the treasury desires to capture the variable funding principal and its reinvestment risk, it will need a funding charge measure, a tool to calculate it, and implementation policies and procedures well vetted.

**Example 4.2.4.** Let's take a look at the PDE with non-zero haircut in section 3.4. Haircut h and the repo rate are both market observable but there is presently no funding market for the residual funding amount $N_t = h\Delta_t S_t$. With h=25%, this amount is sizable. Per discussion above, this funding cost can not enter FVA and shall be handled as a funding charge. The treasury has to come up with the rate charged $r_N$. The funding charge can be computed, for instance,

$$TFC(t) = \frac{1}{\beta_t Q_t} \int_t^T \beta_u Q_u (r_N - r)(hS_t \frac{\partial V}{\partial S}) du$$

where again Q is the joint survival probability, β is the risk free discount factor.

Burgard and Kjaer (2013) employ multiple funding strategies to have open windfall or shortfall upon a dealer's own default. FVA is defined as sum of DVA and FCA (funding cost adjustment) which links to hedging error of the derivative trader's hedging strategy. The derivative asset price would not only depend on the dealer's funding cost, but also dealer's trading desks' strategies. This setup is more in sync as a funding charge measure, rather than a fair value adjustment.

Similarly, Brigo et al, Crepey, and other others have indeed contributed to models to measure the TFC. In fact, all private valuation can be thought of fair value with TFC deducted.

**4.3 Funding cost contributes to bid-ask spreads**

FVA and TFC are complimentary in that FVA is a part of the fair value while TFC contributes to the bid/ask spread off the fair value. Lou (2014) demonstrates that asymmetric funding cost and stock financing cost could contribute significantly to option market making. For exchange traded longer term options, the contribution alone is at the magnitude of the observed bid-ask spreads.

For non-central cleared bilateral OTC derivatives, market depth needs to be carefully examined. In particular, an uncollateralized bilateral trade with non-vanilla features is best treated as an illiquid trade. It could take days to value and negotiation to agree on an unwinding price. Costs from all sources will be considered, funding cost, carrying cost, capital usage, hedging cost or hedge unwinding cost, legal and other fees, etc. A desk has to be prepared to determine its bid and ask, i.e., a private value range.

On a high level, the bid and ask can be caused by asymmetric treasury operation, i.e., borrowing cash from treasury incurs higher rates than depositing cash with treasury. Lou's setup (2010, 2014) is a typical yet extreme case where deposit is at the risk free rate while borrowing at the firm's senior unsecured rate. To long a derivative asset, the desk has the treasury finance the purchase, i.e., borrows cash from the treasury, and to short, she gets paid cash which earns deposit rate. Intuitively she would pay (bid) for less to compensate the cost of financing the purchase and ask for more to compensate the mean return on cash.

As each firm's funding cost is different and treasury operates in its own way, it would be natural to expect derivative desks to maintain their own sets of bid and ask prices. When a customer order arises, the best bid or ask gets hit and that hit becomes a print of the trade and the market price. If customer's order is large, a syndicate could be formed and the final offer is cut with knowledge of market depth. Syndication is in fact how primary debt issuances are getting done and a firm's cash financing rate is determined. The liability side pricing principle implicitly assumes a market mechanism for the derivative liability that resembles the cash debt syndication process and stipulates that the bid or ask priced comparable to the liability side's debt rate is the fair market level.

**4.4 Multi-curve and discounting**

Discounting fully collateralized derivatives at OIS rather than usual LIBOR constitutes the first significant advancement of post-crisis finance. Multi-curve setting separating discounting curve from interest rate indices immediately becomes the new market standard in swaps and interest rate derivatives markets. For uncollateralized trades, FVA is supposed to be the thing, but never quite gets there due to its many definitions and lack of consensus thereof. For example, it is never clear from the FVA literature as how discounting should be done in an uncollateralized bilateral derivative trade. Most resort to the usual risk-free discounting while qualifying payoffs or exposures with different default behaviors, so that there is no clean identification of a discounting curve like the OIS vs LIBOR discounting. The liability-side pricing principle is the first to have a clear answer: uncollateralized payoffs and cashflows are to be discounted locally at the liability-side's financing curve.

With the issue of discounting uncollateralized derivatives settled, models integrating counterparty credit risk and market risk can be developed for various derivatives products. Counterparty risk sensitive fair value can now come out from a single model, instead of a market model for the counterparty risk free price plus separate adjustment calculations. Noticeably, in the first consultative document of BASEL Committee on Banking Supervision on the fundamental review of the trading book (May 2012), there was an agenda of integrating counterparty credit risk with market risk so that there would be no need for a separate CVA risk capital charge. In the second consultative document, however, the committee deemed, for the time being, such an integrated model not readily available and feared its complexity and model risk, and subsequently decided to keep the controversial CVA risk capital charge as a standalone capital charge. The liability-side pricing principle however does offer a simple and intuitive framework to fully integrate counterparty credit risk and market risk and it could potentially reopen the dialogue.

A funding curve is a discounting curve. A popular CVA calculation method is to discount projected and aggregated future cash flow with the cash curve and the risk-free curve separately, the resulting npv difference then being taken as CVA. There are couple variations possible as when to choose which party's curve (EY 2014). For example one might apply counterparty's credit curve to discount if the future net cash flow is incoming or positive, or own curve if outgoing or negative. These types of discounting schemes can best be presented as approximate. As pointed out earlier, the extended PDE exhibits a switching, localized discounting rate dependent on the fair value instead of the net cash flow.

In the liability-side discounting scheme leading to fair value and related valuation adjustments, there is no role for Libor. The funding basis is measured based on the risk free rate or OIS curve, although industry convention typically compares a zero-volatility spread to Libor with the CDS spread of the same tenor. At the time when OIS was proposed to be the discount curve for collateralized trades, most thought Libor could remain its role for uncollateralized trades. This can be partially justified, if we assume the funding basis of the counterparties are same as the basis of Libor to OIS. Specifically, for a derivative book of uncollateralized trades that already have CVA/DVA calculated and use Libor as the discount curve, there is no need to calculate FVA, on a first order approximation basis.

**4.5 Sample numerical results**

Under a simple case where the dealer buys an option from party C, assuming all rates are flat, CVA from B&K denoted as CVA_BK is,

$$CVA\_BK = (1 - e^{-(\lambda_c(1-R_c) + \lambda_B(1-R_B))*(T-t)})V^*(t),$$

To highlight the difference of B&K from the coherent CVA, assuming zero basis, our CVA is given by

$$CVA\_LSP = (1 - e^{-\lambda_c(1-R_c)*(T-t)})V^*(t).$$

Obviously CVA_BK depends on the hazard rate of the dealer, $\lambda_B$ but CVA_LSP does not. The ratios of CVA to the risk-free option price are plotted in Figure 1 with *T-t=5* years, and both recovery rates at 40%.

Because the governing PDE or the risk-neutral pricing formula has an effective discount rate switching on the fair value itself, the fair value solution is in general recursive and would require numerical schemes to address the switching discount rate. A finite difference scheme where an iterative procedure is used to determine the asset/liability boundary (Lou 2015) is adopted to solve the extended BSM PDE under the liability pricing principle.

To demonstrate, we price a shifted stock forward, i.e., an option trade that longs a 45 strike European call and short a 55 strike put of same one year expiry, with the spot price at 50. This trade has positive payoff when stock price $S_T$ is above 50 and negative otherwise, so it involves true bilateral valuation adjustments.

At 50% volatility, 5% risk-free rate and 50 bps repo or stock borrowing cost, the risk-free price is 1.6009. Set bank B's zero recovery CDS at flat 50 bps and liquidity basis 20 bps, and C's CDS at 300 bps, liquidity basis 50 bps, the fair price of the trade is 1.3577. The total valuation adjustment of 0.2432 is decomposed into CVA=0.2501, DVA=0.0342, CFA=0.0410, and lastly DFA=0.0136.

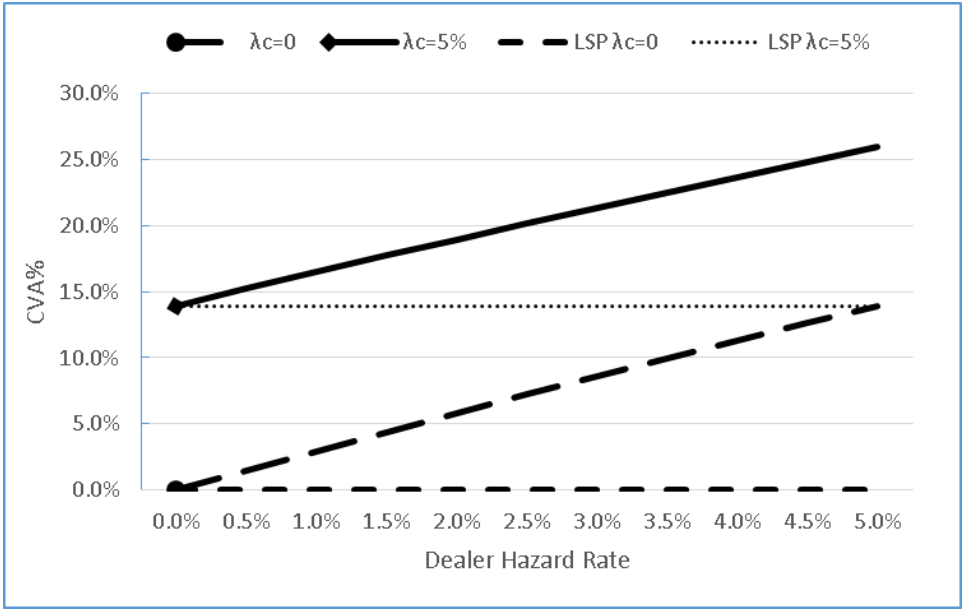

Figure 1. Comparisons of LSP CVA ratio to B&K CVA ratio vs dealer's hazard rate ($\lambda_B$) under two counterparty hazard rates $\lambda_c = 0$ and $\lambda_c = 5\%$.

Obviously, finite difference schemes are of limited uses for CVA and FVA of a large derivative portfolio or netting set. A Monte Carlo simulation procedure would have to be developed.

**5. Conclusion**

Starting from an economic analysis of an uncollateralized option, we find that the liability side party can fund and credit hedge the other party by making a deposit that earns its debt interest rate. An option remains replicable and the newly derived PDE applies counterparty's curve to net derivative receivable and own curve to net payable. In an extension to the risk-neutral pricing formula, the fair value is found to be the expectation of liability-side discounted payoffs. Discounting of bilateral defaultable derivatives is now clearly identified with the liability side's curve, enabling future development of integrated counterparty credit risk and market risk models.

The extended PDE serves to define the total counterparty risk adjustment precisely by discounting at the local liability curve the product of the risk-free price and the credit spread. The adjustment can be broken coherently into CVA -- a default risk component corresponding to CDS, and FVA -- a component due to funding or liquidity basis. Specifically, the same PDE is solved under the synthetic or CDS implied funding curve to arrive at the risky price, then the differences between the risk-free price and the risky price become CVA and DVA, while the differences between the fair price and the risky price become CFA (credit funding adjustment) and DFA (debit funding adjustment). CFA and DFA are collectively called FVA. These measures are symmetric.

This new treatment resolves important FVA debate issues: it is in line with the IFRS 13 standard on fair value measurement; there is no overlap or double counting between DVA and FVA; a derivative asset's fair value will not depend on the holding party's funding cost, i.e., there is no violation of the law of one price; cashflow aggregation and discounting schemes are not accurate. For a funding cost not observable or not market prevailing, it is better to leave it out of FVA and treat it in an internal or treasury funding charge that would impact the desk's bid and offer prices.

**References**

BASEL Committee on Banking Supervision, Consultative Document, Fundamental review of the trading book: A revised market risk framework, May 2012, and October 2013.
Brigo, D. and M. Morini, Closeout Convention Tensions, Risk, December 2011.
Brigo, D., Q. Liu, A. Pallavicini, D. Sloth, Nonlinear Valuation under Collateral, Credit Risk and Funding Costs: A Numerical Case Study Extending Black-Scholes, Handbook in Fixed-Income Securities, Wiley, 2014.
Burgard, C. and M. Kjaer, Partial Differential Equation Representations of Derivatives with bilateral counterparty risk and funding costs, J of Credit Risk, Vol 7, No. 3, Fall, 2011.
Burgard, C. and M. Kjaer, Funding strategies, funding costs, Risk Magazine, December, 2013.
Crepey, S., Bilateral counterparty risk under funding constraints – part I: pricing, Mathematical Finance, part II: CVA, Mathematical Finance, doi: 10.1111/mafi.12005, Dec 2012.
EY, Credit Valuation Adjustment for Derivative Contracts, April 2014.
Gregory, J., Counterparty Credit Risk, Wiley, 2010.
Hull, J. and A. White, The FVA Debate, Risk Magazine, July 2012.
Hull, J. and A. White, Valuing Derivatives: Funding Value Adjustments and Fair Value, Financial Analysts Journal, Vol. 70(3), May/June, 2014.
Longstaff, F., S. Mithal, and E Neis, Corporate Yield Spreads: Default Risk or Liquidity? New Evidence from the Credit Default Swap Market, J of Finance, Vol 60(5), pp 2213-2253, Oct 2005.
ISDA, CSA close-out amount protocol, 2009.
KPMG, FVA – Putting Funding into the Equation, December 2013.
Lou, Wujiang, On Asymmetric Funding of Swaps and Derivatives - A Funding Cost Explanation of Negative Swap Spreads (Nov 5, 2009). Available at SSRN: http://ssrn.com/abstract=1610338.
Lou, Wujiang, Endogenous Recovery and Replication of A Segregated Derivatives Economy with Counterparty Credit, Collateral, and Market Funding -- PDE and Valuation Adjustment (FVA, DVA, CVA) (March 21, 2013). Available at SSRN:http://ssrn.com/abstract=2200249 or http://dx.doi.org/10.2139/ssrn2200249

Lou, Wujiang, Funding in option pricing: the Black-Scholes framework extended, Risk, April, 2015.

Morini, M. and A. Prampolini, Risky Funding: A Unified Framework for Counterparty and Liquidity Charges, Risk Magazine, March , 2011.

Piterbarg, V., Funding beyond discounting: collateral agreements and derivatives pricing, Risk, Feb 2010. pp 97-102.

Tang, Y. and B Li, Quantitative Analysis, Derivatives Modeling, and Trading Strategies in the presence of counterparty credit risk for the fixed income market, World Scientific Publishing, 2007.
21